\documentclass[sigconf,noacm,screen,prologue,dvipsnames]{acmart}

\usepackage[inline]{enumitem}
\usepackage{balance}
\usepackage{microtype}

\usepackage{multicol}

\usepackage{algorithm}
\usepackage{algpseudocode}

\usepackage{tikz, stfloats}
\newcommand*\circled[1]{\tikz[baseline=(char.base)]{
        \node[shape=circle,draw,inner sep=1pt, Maroon, fill=Maroon] (char)
              {\color{white}\scriptsize\textbf{#1}};}%
        }

\usepackage{nicefrac}

\usepackage[]{cleveref}
\crefname{appsec}{Appendix}{Appendices}
\crefformat{section}{\S#2#1#3}
\crefrangeformat{section}{\S#3#1#4--\S#5#2#6}
\crefformat{subsection}{\S#2#1#3}
\crefformat{subsubsection}{\S#2#1#3}
\crefformat{equation}{(#2#1#3)}
\crefrangeformat{equation}{(#3#1#4--#5#2#6)}
\crefmultiformat{equation}{(#2#1#3)}{ and~(#2#1#3)}{, (#2#1#3)}{ and~(#2#1#3)}
\crefformat{figure}{Fig.~#2#1#3}
\crefrangeformat{figure}{Figs. #3#1#4--#5#2#6}
\crefmultiformat{figure}{Figs.~#2#1#3}{ and~#2#1#3}{, #2#1#3}{ and~#2#1#3}
\crefformat{algorithm}{Alg.~#2#1#3}
\crefformat{table}{Table~#2#1#3}
\crefrangeformat{table}{Tables~#3#1#4--#5#2#6}
\crefmultiformat{table}{Tables~#2#1#3}{ and~#2#1#3}{, #2#1#3}{ and~#2#1#3}

\usepackage[compact]{titlesec}
\titlespacing*{\section}{0pt}{6pt plus 4pt minus 2pt}{2pt plus 2pt minus 2pt}
\titlespacing*{\subsection}{0pt}{4pt plus 2pt minus 1pt}{2pt plus 1pt minus 1pt}
\titlespacing*{\subsubsection}{0pt}{4pt plus 2pt minus 1pt}{2pt plus 1pt minus 1pt}

\setcopyright{none}
\renewcommand\footnotetextcopyrightpermission[1]{} 

\title{BayesPerf: Minimizing Performance Monitoring Errors Using Bayesian Statistics}

\author{Subho S. Banerjee}
\orcid{0000-0001-7187-6569}
\affiliation{%
    \institution{University of Illinois at Urbana-Champaign}
    \city{Urbana}
    \state{Illinois}
    \country{USA}}
\email{ssbaner2@illinois.edu}

\author{Saurabh Jha}
\affiliation{%
    \institution{University of Illinois at Urbana-Champaign}
    \city{Urbana}
    \state{Illinois}
    \country{USA}}
\email{sjha8@illinois.edu}

\author{Zbigniew T. Kalbarczyk}
\affiliation{%
    \institution{University of Illinois at Urbana-Champaign}
    \city{Urbana}
    \state{Illinois}
    \country{USA}}
\email{kalbarcz@illinois.edu}

\author{Ravishankar K. Iyer}
\orcid{0000-0003-2245-3038}
\affiliation{%
    \institution{University of Illinois at Urbana-Champaign}
    \city{Urbana}
    \state{Illinois}
    \country{USA}}
\email{rkiyer@illinois.edu}

\begin{document}
    \begin{abstract}
Hardware performance counters (HPCs) that measure low-level architectural and microarchitectural events provide dynamic contextual information about the state of the system.
However, HPC measurements are error-prone due to non determinism (e.g., undercounting due to event multiplexing, or OS interrupt-handling behaviors).
In this paper, we present BayesPerf, a system for quantifying uncertainty in HPC measurements by using a domain-driven Bayesian model that captures microarchitectural relationships between HPCs to jointly infer their values as probability distributions.
We provide the design and implementation of an accelerator that allows for low-latency and low-power inference of the BayesPerf model for \texttt{x86} and \texttt{ppc64} CPUs.
BayesPerf reduces the average error in HPC measurements from 40.1\% to 7.6\% when events are being multiplexed.
The value of BayesPerf in real-time decision-making is illustrated with a simple example of scheduling of PCIe transfers.
\end{abstract}



    \maketitle
    \pagestyle{empty}
    \section{Introduction}
\label{sec:intro}

Hardware performance counters (HPCs) are widely used in profiling applications to characterize and find bottlenecks in application performance.
Even though HPCs can count hundreds of different types of architectural and microarchitectural events, they are limited because those events are collected (i.e., multiplexed) on a fixed number of hardware registers (usually 4--10 per core).
As a result, they are error prone because of application, sampling, and asynchronous collection behaviors borne out of multiplexing.
Such behavior in HPC measurements is not a new problem, and has been known for the better part of a decade~\cite{Ammons1997,Zellweger2016,Weaver2008,Mytkowicz2007, Dimakopoulou2016,Lv2018,Weaver2013}.

\textbf{Targeted Need.}
Traditional approaches of tackling HPC errors have relied on collecting measurements across several application runs, and then performing offline computations to
\begin{enumerate*}[label=(\roman*)]
    \item impute missing or errored measurements with new values (e.g.,~\cite{Weaver2008}); or
    \item dropping outlier values to reduce overall error (e.g.,~\cite{Lv2018}).
\end{enumerate*}
Both of these require time and compute resources for collecting training data and inference, thus are suitable for offline analysis (like profiling).
These techniques are untenable in emergent applications that use HPCs as inputs to complete a feedback loop and make dynamic real-time decisions that affect system resources using a variety of machine learning (ML) methods.
Examples include online performance hotspot identification (e.g.,~\cite{Gan2019}), userspace or runtime-level scheduling (e.g.,~\cite{Delimitrou2013, Banerjee_Symphony, Zellweger2016, Giceva2014, Baumann2009}), and power and energy management (e.g.,~\cite{Pothukuchi2018, Pothukuchi2019, Tarsa2019, Ding2019}), as well as attack detectors and system integrity monitors~\cite{Das2019}.
In such cases, the HPC measurement errors propagate, get exaggerated, and can lead to \textit{longer training time} and \textit{poor decision quality}  (as illustrated in \cref{sec:scheduling}).
This is not surprising because ML systems are known to be sensitive to small changes in their inputs (e.g., in adversarial ML)~\cite{NIPS2014_5423, 10.1145/3359789.3359847, jha2020ml}.
As we will show in \cref{sec:background}, HPC measurement errors can be large (as much as 58\%); hence they must be explicitly handled. 

This paper presents BayesPerf, a system for quantifying uncertainty and correcting errors in HPC measurements using a domain-driven Bayesian model that captures micro-architectural relationships between HPCs.
BayesPerf corrects HPC measurement errors at the system (i.e., CPU and OS) level, thereby allowing the down-stream control and decisions models that use HPCs to be simpler, faster and use less training data (if used with ML).
The proposed model is based on the insight that even though individual HPC measurements might be in error, groups of different HPC measurements that are related to one another can be jointly considered---to reduce the measurement errors---using the underlying statistical relationships between the HPC measurements.
We derive such relationships by using design and implementation knowledge of the microarchitectural resources provided by CPU vendors~\cite{IntelSDM,IBM2017_Perf}.
For example, the number of LLC misses, the size of DMA transactions, and the DRAM bandwidth utilization are related quantities,\footnote{In a simple processor, DRAM Bandwidth = (LLC misses \texttimes{} Cache line size+ \# DMA Transactions \texttimes{} Transaction size)/Clocks.} and can be used to reduce measurement errors in each other.

\textbf{Approach \& Contributions.}
The key contributions are:
\begin{enumerate}[noitemsep,nolistsep,leftmargin=*]
    \item \emph{The BayesPerf ML Model.} We present a probabilistic ML model that incorporates microarchitectural relationships to combine measurements from several noisy HPCs to infer their true values, as well as quantify the uncertainty in the inferred value due to noise. Hence allowing:
    \begin{enumerate}[noitemsep,nolistsep]
        \item improving decision-making with explicit quantification of HPC measurement uncertainty.
        \item reduced need for aggressive (high-frequency) HPC sampling (which negatively impacts application performance) to capture high-fidelity measurements, thereby increasing our observability into the system.
    \end{enumerate}
    \item \emph{The BayesPerf Accelerator.} To enable the use of BayesPerf ML model in latency-critical, real-time decision-making tasks, this paper presents the design and implementation of an accelerator for Monte Carlo-based training and inference of the BayesPerf model.
    The accelerator exploits 
    \begin{enumerate}[noitemsep,nolistsep]
        \item high-throughput random-number generators.
        \item maximal parallelism based on the statistical relationships mentioned above, to rapidly sample multiple parts of the BayesPerf model in parallel.
    \end{enumerate}
    \item \emph{A Prototype Implementation.} We describe an FPGA-based prototype implementation of the BayesPerf system (on a Xilinx Virtex 7 FPGA) for Linux running on Intel x86\_64 (Sky Lake) and IBM ppc64 (Power9) processors.
    The BayesPerf system is designed to provide API-compatibility with Linux's \texttt{perf} subsystem~\cite{Perf}, allowing it to be used by \emph{any} userspace performance monitoring tool for both \texttt{x86\_64} and \texttt{ppc64} systems.
    Our experiments demonstrated that BayesPerf reduces the average error in HPC measurements from 40.1\% to 7.6\% when events are being multiplexed, which is an overall 5.28\texttimes{} error reduction.
    Further, the BayesPerf accelerator provides an 11.8\texttimes{} reduction in power consumption, while adding less than 2\% read latency overhead over native HPC sampling.
    \item \emph{Increasing training and model efficiency of decision-making tasks.} We demonstrate the generality of the BayesPerf system by integrating it with a high-level ML-based IO scheduler that controls transfers over a PCIe interconnect.
	We observed that the training time for the scheduler was reduced by $37\%$ (\textasciitilde 52 hr reduction) and the average makespan of scheduled workloads decreased by 19\%.
\end{enumerate}

The remainder of the paper is organized as follows.
First in \cref{sec:background}, we discuss the sources of HPC measurement errors.
Then in \cref{sec:overview} we provide an overview of the design of the BayesPerf system.
\cref{sec:ml_model} describes the formulation, training and inference of the ML model used to correct errors.
\cref{sec:accelerator} describes the accelerator that allows inference on the ML model in real-time. Then in \cref{sec:eval} we discuss a prototype implementation and it's evaluation. 
Finally, in \cref{sec:related_work} and \cref{sec:conclusion}, we put BayesPerf in perspective of traditional methods, and describe future challenges, respectively.

    \section{Background: HPC Errors}
\label{sec:background}

Every modern processor has a logical unit called the Performance Monitoring Unit (PMU), which consists of a set of HPCs.
An HPC counts how many times a certain event occurs during a time interval of a program's execution.
The number and configurability of the HPCs vary across processor vendors and microarchitectures.
For example, modern Intel processors have three fixed HPCs (which measure ISA-related events) and eight programmable HPCs per core (which measure microarchitectural events and are split between the SMT threads on the core)~\cite{IntelAORM}.
The events measured by an HPC are vendor-specific and microarchitecture-dependent, and vary with processor models within the same microarchitecture.
For example, an Intel Haswell CPU has 400 programmable events, compared to the 1623 events on a HaswellX CPU; both have the same number of HPC registers per core (three + eight)~\cite{Zellweger2016}.
Therefore, one must carefully pick and configure which events to monitor with the available registers.

\begin{figure}[!t]
    \centering
    \includegraphics[width=\columnwidth]{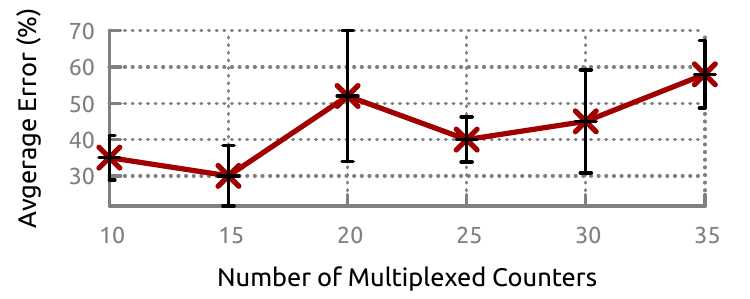}
    \caption{Errors due to event multiplexing in HPC measurements across ten application runs.}
    \label{fig:hpc_errors}
\end{figure}

\textbf{Reading HPCs.}
Performance counters can be read using:
\begin{enumerate}[noitemsep,nolistsep,leftmargin=*]
    \item \emph{Polling:} The HPCs can be read at any instant by using specific instructions to write (to configure the HPC) and read (to poll the value of an HPC) model-specific registers (MSRs) that represent HPCs.
    For example, \texttt{x86\_64} uses specific instructions to read (i.e., \texttt{rdmsr}) from and write (i.e., \texttt{wrmsr}) to MSRs, respectively; both instructions require OS-level access privilege, and hence are performed by the OS on behalf of a user.
    Here, one HPC is programmed to count only one event during the execution of a program.
    Hence, polling is ineffective, as the number of events that can be simultaneously measured is limited by the number of available hardware registers.
    \item \emph{Sampling:} HPCs also support sampling of counters based on the occurrence of events, thereby letting multiple events timeshare a single HPC~\cite{May2001, Mytkowicz2007}.
    This feature is enabled through a specific interrupt, called the Performance Monitoring Interrupt (PMI), which can be generated after the occurrence of a certain number of events (i.e., a predetermined threshold).
    The interrupt handler then polls (i.e., samples) the HPC.
    The multiplexing of events occurs through a separate scheduling interrupt that is triggered periodically to change the configuration of the HPCs and swap events in and out.
    The collected measurements are generally scaled to account for the time they were not scheduled to a HPC~\cite{Dimakopoulou2016}, and that can lead to making erroneous measurements.
    Sampling is necessary due to the severe disparity between the numbers of events types and the number of counters.
\end{enumerate}

\textbf{Sources of Errors.} 
In addition to the errors due to event multiplexing, HPCs demonstrate other modalities of measurement error. 
For example, HPC measurements can vary across runs because of OS activity, scheduling of programs in multitasking environments, memory-layout, and memory-pressure, and varied multi-processor interactions may change between different runs.
Nondeterminism in OS behavior (e.g., servicing of hardware interrupts) also plays a significant role in HPC measurement errors~\cite{Weaver2013}.
Performance counters have also been shown to over count certain events on some processors~\cite{Weaver2013}.
Finally, the implementation of userspace and OS-kernel-level tools can cause different tools to provide different measurements for the same HPCs in strictly controlled environments for the same application.
The variations in measurements may result from the techniques involved in acquiring them, e.g., the point at which they start the counters, the reading technique (polling or sampling), the measurement level (thread, process, core, multiple cores), and the noise-filtering approach used.

\begin{figure*}[!t]
    \centering
    \includegraphics[width=\textwidth]{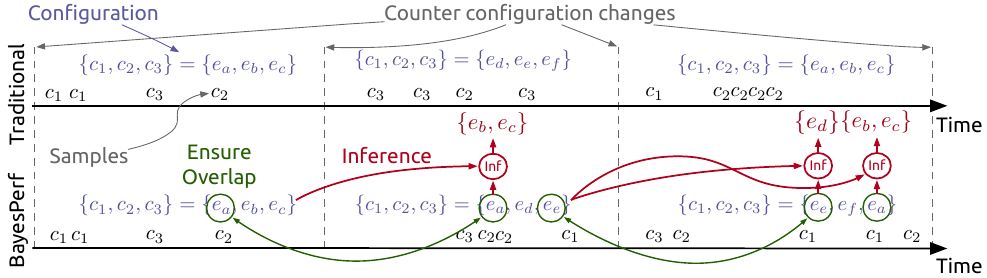}
    \caption{Overview of the BayesPerf ML model.}
    \label{fig:bayes_perf_overview}
\end{figure*}

\textbf{Measurement Errors.}
As a result of this non-determinism, quantifying error in HPCs is difficult as there is no way to get ``ground truth'' measurements because of inherent variations in measurements.
In this paper, we define HPC error as magnitude of difference between corresponding HPC measurements made in two runs of a workload, one in polling and other in sampling mode.
The correspondence between the two HPC traces (time-series) is established by \textit{dynamic time warping}~\cite{Berndt1994} that calculates an ``alignment'' between the two time series datasets using edit-distance.\footnote{This definition of error is based on prior work on HPC errors~\cite{Lv2018}.}

\cref{fig:hpc_errors} illustrates the net effect of measurement errors on the fidelity of an HPC counter using Linux's \texttt{perf} subsystem.
In this case, the baseline dataset is collected using polling, and the target dataset is collected using sampling, each on 10 independent application runs capturing both variations in a single run, and variations across runs.
We observe a $58 \pm 9.3\%$ average error in HPC measurements when 35 on-core events are being multiplexed on an Intel processor, compared to the baseline of polling 4 events at a time.\footnote{The experimental setup is described in detail in \cref{sec:setup}.}

\textbf{Errors in Derived Events.}
Such high error is particularly troubling, as it is quite conceivable to count 35 events simultaneously, particularly for measuring \textit{derived events}.
Derived events are obtained by combining individual HPC measurements in a mathematical expression.
Consider for example, the ``\texttt{Backend\_Bound\_SMT}'' derived event on Intel BroadwellX processor. It measures the fraction of \textmu{}ops issue slots utilized in a core, and alone takes measurements from 16 HPCs to compute~\cite{IntelSDM}.
This information might be valuable in a OS-level scheduler that controls an SMT processor, with the objective of minimizing interference between CPU-bound processes/threads.
Often such information would be conflated with other derived metrics like ``\texttt{Memory\_Bound}'' and ``\texttt{Frontend\_Bound\_SMT}'', which together would require the use of 29 unique counters.
That according to \cref{fig:hpc_errors} would incur an average error of \textasciitilde 45\%.
This is further exasperated by the fact that the HPCs need to be counted per-SMT thread, per-core, and per-socket. 
For example, in an average 2-socket server system this would imply collecting thousands of counters (i.e., 2784 HPCs = 29 counters \texttimes{} 24 cores \texttimes{} 2 sockets).

\textbf{Adding More Registers?} 
A relevant question to ask is whether the HPC-error problem will disappear if more HPC registers are added into future CPUs. 
The short answer is that it will not, because as we continue to add more monitors, the system complexity increases which is untenable in commercial CPUs that are often driven by other practical considerations.
Hence, HPC counters will eventually always end up introducing the sampling-based error.

    \section{Approach Overview}
\label{sec:overview}

\textbf{Key Insight.}
The key insight that drives this work is that microarchitectural invariants (e.g.~\cite{IntelSDM,IBM2017_Perf,Kernel_perf}) can be applied to measured HPC data to estimate whether it is, in fact, in error (i.e., a detector).
Further, we can quantify the ``uncertainty'' of an HPC measurement by quantifying the probability of deviation from that invariant (i.e., its egregiousness).
When the above is applied to a group of HPC measurements, each targeting different microarchitectural units, the underlying invariants can be composed, encoded as statistical relationships, i.e., joint probability distributions, which can then be composed into larger \emph{probabilistic graphical models}.
We then use a Bayesian inference approach to integrate the data and prior knowledge of the system to effectively attenuate the high error measurements and significantly amplify correct measurements, all in real-time.
This works in practice as the number of HPCs with lower errors are generally more numerous than those with higher error (also verified by our observations), hence they bias the aggregate results to the lower errored values.
As a result, BayesPerf significantly outperforms traditional purely data-driven statistical approaches for outlier detection.

\textbf{BayesPerf ML Model.}
Below, we provide a high-level description of the model, using the example illustrated in \cref{fig:bayes_perf_overview}. 
In this example, the goal is to measure (by multiplexing) a set of events $\{e_a, \dots, e_f\}$, on a set of HPCs $\{c_1, c_2, c_3\}$.

\emph{Deciding Schedules of HPCs:} 
BayesPerf first determines a schedule of how the events are multiplexed on the HPCs.
The schedule consists of a set of \textit{HPC configurations} that are collected over time.
We define an HPC configuration as a mapping between counters and events, that defines which counters are collected at an instant of time.
The notation $\{c_1, c_2, c_3\} = \{e_a, e_b, e_c\}$ is used to define such a configuration, and imply that $c_1$ counts $e_a$.
The scheduling process is driven primarily by the microarchitectural considerations of the available HPCs and the types of events that each one can measure, i.e., as not all HPCs can measure all events.
Traditional HPC measurement tools, like the Linux \texttt{perf} subsystem trigger HPC configuration changes in a round-robin manner, based on a periodic hardware timer-driven interrupt (see \cref{fig:bayes_perf_overview}).
BayesPerf uses a similar interrupt driven approach, but does not use round-robin to build a schedule of configurations.
\textit{It creates configurations of overlapping counters, such that each set of counters have ``statistical relationships'' to other events in preceding and subsequently scheduled configurations.}
For example, in \cref{fig:bayes_perf_overview}, $e_a$ and $e_e$ are such overlapping events.
As we will show in \cref{sec:ml_model}, these ``statistical relationships'' can be derived based on microarchitectural invariants (i.e., domain knowledge) that tie together the resources underlying the measurements.
BayesPerf encodes those invariants as generative \emph{joint-} and \emph{conditional-probability distributions} for the processors used in our experiments.
    
\emph{Inferring Unscheduled Events:} At each instant of time, BayesPerf then uses sampled data from the overlapping events to compute a full posterior distribution (i.e., the likely values and their associated uncertainties) of the unscheduled events using a Bayesian inference approach.
Consider $e_b$ in the second time slice of \cref{fig:bayes_perf_overview}.
It is calculated using its' own samples from the previous time slice and the samples of $e_a$ (which is the event repeated across time slice one and two) in the current time slice.
The result of the Bayesian inference using the sampled data is a probability distribution $\Pr(e_b^t| e_b^{t-1}, e_a^t)$ at time $t$; this distribution not only gives us an estimate of $e_b$ (i.e., by finding the most likely value of $e_b$ under the distribution), but also quantifies uncertainty (i.e., using the probability value $\Pr(e_b| \dots)$) in that estimate.
The compositional nature of Bayesian inference allows chain events across multiple time slices, if the overall set of events to be measured is large, albeit at the cost of larger uncertainty in the estimate. 
For example, in \cref{fig:bayes_perf_overview} the chain of events $(e_b \rightarrow e_a) \rightsquigarrow (e_a \rightarrow e_e) \rightsquigarrow (e_e \rightarrow e_d)$ can be used directly estimate $e_b$ from samples of $e_a$, but also transitively estimate it from samples of $e_e$. 
Here ``$\rightarrow$'' describes the above statistical relationships between events in a configuration (i.e., in a single time slice), and ``$\rightsquigarrow$'' describes data collected between overlapping events across time slices.
    
The BayesPerf system then allows an user to poll the posterior probability distributions of any of the events being collected.
These distributions can be passed along (i.e., integrated) into higher-level ML/control frameworks or used directly to compute error bounds of HPC measurements.

\begin{figure}[!t]
    \centering
    \includegraphics[width=\columnwidth]{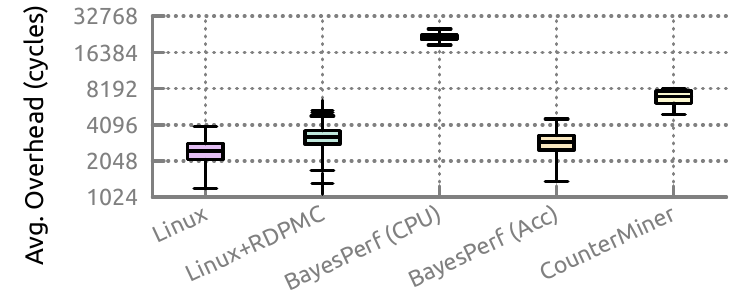}
    \caption{Latency overhead of reading counters with BayesPerf compared to traditional methods on an x86 CPU.}
    \label{fig:overhead}
\end{figure}

\textbf{BayesPerf Accelerator.}
Though the BayesPerf ML model is able to provide significantly higher-quality samples from the raw HPC measurements, it introduces the additional runtime overhead of performing Bayesian inference on every new measurement polled by the user.
Consider \cref{fig:overhead}; it shows the average overhead (over 100 reads) of reading a HPC value using the Linux kernel's (perf subsystem) \texttt{read()} system call (i.e., polling), the x86\_64 \texttt{rdpmc} instruction to read HPCs in userspace, a purely CPU implementation of the BayesPerf ML model (using TensorFlow Probability~\cite{Tran2016, Dillon2017}), an FPGA accelerated version of BayesPerf (described later in \cref{sec:accelerator}), and CounterMiner\cite{Lv2018} (described later in \cref{sec:eval} and used as a baseline in our evaluation).
We observe that a single HPC read when  the CPU implementation of BayesPerf is being used has approximately $9\times$ longer latency than native polling of the HPC.
In order to reduce the latency, we introduce an accelerator that parallelizes the process of computing posterior inference on the BayesPerf ML model.
The accelerator largely builds upon our prior work~\cite{Banerjee_ASPLOS2019} in building MCMC accelerators that treats a lack of statistical dependencies between variables as a scope for parallel execution.
Using the accelerator, BayesPerf adds less than 2\% overhead in read latency compared to the native solution.
Our implementation of the accelerator on a PCIe-attached FPGA device can take advantage of modern cache-coherent accelerator-processor communication protocols like CAPI~\cite{CAPI2015}, and essentially provide users with the same interface and same performance characteristics they could get if they were natively polling the OS for HPC measurements.

    \section{The BayesPerf ML Model}
\label{sec:ml_model}

In this section we first discuss formalization of the HPCs and events for a generic CPU.
Then, in \cref{sec:scheduling_hpcs}, we discuss the problem of scheduling sets of performance counters onto available HPCs.
Finally, in \cref{sec:inference}, we discuss an inference strategy to compute the posterior distribution of a single event based on generated schedule and HPC measurement samples.

\textbf{Formalism.}
We assume that every processor has a pre-determined number of fixed and programmable HPCs.
We refer to them as $n_f$ and $n_p$, respectively.
The HPCs themselves are indexed and referred to as $f_1 \dots f_{n_f}$ for the fixed HPCs and $c_1 \dots c_{n_p}$ for the programmable HPCs.
The processor as a whole has a set $E = \{ e_1, \dots, e_{n_e} \}$ of $n_e$ architectural and microarchitectural events that are measured using $f_*$ and $c_*$.
At any point in time, the programmable HPCs are configured to count any one of the events in $E$.
The instantaneous mapping between counters and events is called a \emph{configuration}.
Fixed HPCs are not considered in a configuration, as they cannot be programmed.
Not all programmable HPCs will be able to count all events (i.e., all configurations might not be valid), depending on microarchitectural and implementation considerations. 
For example, an Intel off-core response event requires one HPC and one MSR register, and the \texttt{L1D\_PEND\_MISS.PENDING} event can be only counted on the third HPC on Haswell/Broadwell systems. 
Configuration validity constraints are known ahead of time, can be dynamically checked, and must always be satisfied.
BayesPerf uses the Linux's builtin validity checker.

A sample $s_j$ is generated from an HPC $c_i$ (i.e., an interrupt is fired to read the value of a counter and store it in memory) when a particular threshold $\tau_{i, k}$ is reached on one of the fixed HPCs $f_k$.\footnote{In general, this triggering event occurs based on the number of clock cycles or number of instructions executed.}
That process is denoted by $s_j \sim c_i$ if $f_k \geq \tau_{i,k}$.
In addition to the value of the counter, the sampling process also records two time measurements, $t_r^i$ and $t_e^i$, where $t_e^i \leq t_r^i$.
They correspond to the total time the application has been running, and the total time for which an event has been sampled (i.e., it has been enabled), respectively.
Traditional approaches (e.g., one that is used in Linux) use these times to correct HPC undercounting errors and assume that the true value of a performance counter is scaled according to $s_j \mapsto s_j \times \nicefrac{t_r^i}{t_e^i}$.

\textbf{Statistical Dependencies.}
Some subsets of events in $E$ have statistical relationships between them.
Those statistical relationships are described by \emph{joint probability distribution functions}.
For example, if $e_1$ and $e_2$ share such a relationship, then it is represented by their joint probability distribution $\Pr(e_1, e_2; \Theta)$. 
Where, $\Theta$ refers to all tunable or learnable parameters of the distribution.

We assert that if nothing is known about the statistical relationships between the events, then $\Pr(e_i, \dots)$ can be approximated by a neural network and trained using data from HPCs.
However, for most real systems, knowledge about the underlying microarchitectural resources being counted in a HPC can be correlated together to describe $\Pr(e_i, \dots)$.
To do so, we use algebraic models of the composition of HPC measurements by using information about the CPU microarchitecture found in processor performance manuals~\cite{Yasin2014,IntelSDM,IBM2017_Perf}.
For example, in an Intel x86 Sandy Bridge microarchitecture~\cite{TMAM, Yasin2014}, the fraction of cycles a CPU is stalled because of DRAM access is given by 
$(1 - \text{\texttt{Mem\_L3\_Hit\_Frac}}) \times \nicefrac{\text{\texttt{STALLS\_L2\_PENDING}}}{\text{\texttt{CLKS}}}$.
Those stalls can be caused by either DRAM bandwidth issues or DRAM latency issues, which in turn can be measured as
$\nicefrac{\text{\texttt{ORO\_DRD\_BW\_Cycles}}}{\text{\texttt{CLKS}}}$, and 
$\nicefrac{\text{\texttt{ORO\_DRD\_Any\_Cycles}}}{\text{\texttt{CLKS}}} - \nicefrac{\text{\texttt{ORO\_DRD\_BW\_Cycles}}}{\text{\texttt{CLKS}}}$ respectively.
Here, \texttt{ORO\_DRD\_Any\_Cycles}, \texttt{ORO\_DRD\_BW\_Cycles}, \texttt{Mem\_L3\_Hit\_Frac}, \texttt{STALLS\_L2\_PENDING}, and \texttt{CLKS} correspond to a set of fixed and programmable events, which are related to each other via the algebraic relations described above.
Given the equivalence of those three computed quantities, we can compute one, given values of the other.
When some of these events are reported with measurement errors, the equivalence relationship becomes statistical (i.e., capture randomness because of errors).
We then define a distribution function for individual events, where only valid combinations of the event values have a non zero probability of occurrence.

\subsection{Scheduling}
\label{sec:scheduling_hpcs}

\textbf{Problem.}
Given statistical dependencies between events, we need to ensure that the configurations created for two consecutive time slices (i.e., scheduler quanta) have at least one overlapping event in order to establish either a first-order or a transitive statistical relationship between consecutive time slices.
For example, if we have four events $e_1$ to $e_4$ that are related by $f(e_1, e_2)$ and $g(e_2, e_3, e_4)$, we must ensure that samples of $e_2$ occur repeatedly across multiple time slices.
Given (from a profiling application) an original schedule of configurations $C_1 \rightarrow C_2 \rightarrow \dots \rightarrow C_n$, where $C_i$ executes in time slice $i$, into another schedule of $C'_i$s such that transitive statistical relationships hold, such that the validity criteria holds on each $C'_i$.
In the case when it is not possible ensure the validity criteria on every $C'_i$, we break the chain of repeated events, and start over again from a valid configuration.

\textbf{Solution.}
The first step of the scheduling process is to aggregate all the statistical dependencies available for the processor in question into a graphical structure.
The graph is produced by expanding the scheduled chain $C_1 \rightarrow \dots \rightarrow C_n$ using the statistical relationships between the events in the chain.
In the ML/statistics community, such a graph commonly referred to as \emph{probabilistic graphical model}, and more specifically identified as a \emph{factor graph} (FG)~\cite{Koller2009}.
Remember from above that the statistical dependencies between the events are specified as joint probability functions $\Pr(S_i)$,\footnote{We use the shorthand $\Pr_i = \Pr(S_i)$.} where $S_i \subseteq E$.
Using those functions, we generate a bipartite FG 
$G = (E \cup \{\Pr_1, \dots \Pr_n\}, \{(e, \Pr_i) | e \in S_i \forall i\})$.
The FG represents the joint distribution of \textit{all} the events in the schedule, composed together from every individual joint distribution.

Now, given the FG and two consecutive configurations from a schedule $C_t$ and $C_{t+1}$ (with events $E_t, E_{t+1} \subseteq E$ respectively), our scheduling problem reduces to 
\begin{enumerate*}[label=(\roman*)]
    \item finding whether $E_t$ and $E_{t+1}$ share an event such that the transitive statistical dependency is met; and
    \item if they do not share such a dependency, producing the shortest sequence of $C'_*$ such that $C_t \rightarrow C'_{(1)} \rightarrow \dots \rightarrow C_{t+1}$.
\end{enumerate*} 
Solution of the first of the two problems is straightforward.
We do it by computing the \emph{Markov blanket}~\cite{Koller2009} of the sets $E_t$ and $E_{t+1}$ under the factor graph. 
The Markov blanket $B_{x_i}$ of a variable $x_i$ in the factor graph defines a subset of $x_{\lnot i}$ such that $x_i$ is conditionally independent of $x_{\lnot i}$ given $B_{x_i}$.
If the Markov blankets of $E_t$ and $E_{t+1}$ overlap (i.e., $B_{E_t} \cap B_{E_{t+1}} \neq \varnothing$), then we are guaranteed that there exists at least one event that shares transitive dependencies between the time slices.
The second problem is a little more involved.
It can be solved by finding the shortest path (assuming unit cost for each edge traversed) from each $e \in E_t$ to each $e' \in E_{t+1}$ in the FG.
That can be accomplished using Djikstra's algorithm, checking validity of the path at every step.
In addition to the graph traversal, one must also apply the following optimizations to prune unnecessary $C'_*$s.
\begin{enumerate}[noitemsep,nolistsep,leftmargin=*]
    \item \emph{Removing Common Steps:} If an intermediate step $C'_{i}$ exists such that the Markov blankets $B_{e_1}, B_{e_2}, \dots B_{e_n}$ of events $e_1, e_2, \dots e_n$ overlap, the next transition state of the schedule can be condensed.
    That is, if there exists an $e_* \in B_{e_1} \cap \dots B_{e_n}$, then composition of statistical relationships can happen through $e_*$, instead of through the larger set of events, i.e.,
    $C'_{i+1} \mapsto (C'_{i+1} \setminus \{e_1,\dots e_n\})\cup e_*$
    \item \emph{Removing Redundant Steps:} If there exists two steps $C'_{i}$ and $C'_{i+1}$ such that there is no change in the Markov blanket (i.e., $B_{E_{i}} = B_{E_{i+1}}$), then we can skip the transition $C'_{i+1}$ and instead transition to $C'_{i+2}$.
    That situation can occur because the Markov blankets in individual traversals $e \rightsquigarrow e'$ will change at every step; however, the union of all such blankets might not change.
    If it does not change, we have enough statistical information to skip the $i+1$\textsuperscript{th} step and go directly to $i+2$.
\end{enumerate}

\textbf{Checking Validity of the Configuration.}
A key challenge in determining a valid transformation of a schedule is that of identifying the configurations that do not satisfy the microarchitectural constraints placed on HPCs.
We check the validity of a new schedule using Linux's perf\_event subsystem.
It allows us to iterate over all HPCs in a configuration until it reaches an event that it fails to schedule, thereafter notifying the user of validity failure.
To maximize the use of available counters, the perf iteration strategy starts with the most constrained events and goes to the least constrained events in a configuration.
Linux's native scheduling for a group of events happens independently per PMU and per logical core.
As some PMUs are shared between threads of the same core or package, their availability may change depending on what events are being measured on the other cores.

\subsection{Modeling Errors in Event Samples}
The first step to computing the full posterior distribution is to model errors in the capture of samples from HPCs.
Recall that we listed sources of such errors in~\cref{sec:background}.
For a single event $e$ programmed in an HPC $c$, if the error in measurement $e_c$ can be modeled, then the measured/sampled values $m_c$ can be modeled in terms of the true value $v_c$ plus measurement noise $e_c$, i.e., $m_c = v_c + e_c$.
Here, we focus only on random errors, by assuming zero systematic error.
That is a valid assumption because the only reason for systematic errors will be hardware or software bugs. 
We assume that the error can be modeled as $e_c\sim \mathcal{N}(0, \sigma)$ for some unknown variance $\sigma$, hence $\Pr(m_c \mid v_c) = \mathcal{N}(m_c, \sigma)$~\cite{Weaver2008}.
Now, given $N$ samples of HPC, we compute their sample mean $\mu$ and sample variance $S$.
A scaled and shifted Student's \emph{t}-distribution describes the marginal distribution of the unknown mean of a Gaussian, when the dependence on variance has been marginalized out~\cite{Gelman1995}, i.e., $v_c \sim \mu + \nicefrac{S}{\sqrt{N}}~Student(\nu = N-1)$.
In all our experiments, the confidence level of the \emph{t}-distribution was set to 95\%.
Now, since the measurement error model for an HPC is stochastic, when samples from these models are used in the algebraic relationships described above, they too become stochastic in nature.
\textit{The FG becomes one unified graphical representation of all of these statistical relationships, i.e., between the errored samples and true values of events, as well as among different events that measure related aspects of the CPU's microarchitecture.}

\subsection{Inference Strategy}
\label{sec:inference}

Once we have computed a schedule that ensures that events with statistical dependencies between them are measured in consecutive time slices, the next goal is to utilize the measurements to produce a posterior distribution for an event.
Recall \cref{fig:bayes_perf_overview}. In each scheduling time slice, we have measurements/samples from the current slice and the preceding slice.
However, because of the transitive statistical dependencies, we would like to jointly compute inference for the FG (i.e., compute the posterior probability of some event in the FG given the sampled data) for some $k$ time slices into the past.

\begin{algorithm}[!t]
    \caption{General EP algorithm.}
    \label{alg:ep}
    \small
    \begin{algorithmic}[1]
     \renewcommand{\algorithmicrequire}{\textbf{Input:}}
     \renewcommand{\algorithmicensure}{\textbf{Output:}}
        \Require Target distribution $f(\theta) = \prod f_k(\theta)$ 
        \Ensure Global approximation $g(\theta) = \prod g_k(\theta)$
        \State Choose initial $g_k(\theta)$
        \For {$k \in \{0, \dots K-1\} \land$ until $g_k$ converges}
            \State $g_{-k}(\theta) \propto \nicefrac{g(\theta)}{g_k(\theta)}$
            \Comment{Cavity distribution}
            \State $g_{\setminus k}(\theta) \propto \Pr(y_k | \theta) g_{-k}(\theta)$
            \Comment{MCMC}
            \State $g^{new}(\theta) \propto g_{\setminus k}(\theta)$
            \Comment{Local update}
            \State $\Delta g_k(\theta) \approx \nicefrac{g^{new}(\theta)}{g(\theta)}$
            \State $g(\theta) \gets g(\theta) \Delta g_k(\theta)$
            \Comment{Global update}
        \EndFor\\
        \Return $\{ g_k(\theta) | k \in [0, K) \}$
     \end{algorithmic} 
\end{algorithm}

Our approach to performing this computation with low-latency guarantees utilizes the idea that one can break the larger problem into $k$ smaller parts, performing inference on each of the $k$ parts, and then put the results together to get an approximate posterior inference, i.e., similar to map-reduce.
There are two difficulties with such algorithms, as they are usually constructed.
First, each of the $k$ pieces has only partial information; as a result, for any of the pieces, a lot of computation is wasted in places that are contradicted by the other $k-1$ pieces.
Second, the partial results from the $k$ pieces must be carefully combined together to ensure that the prior (which is embedded into the FG model) is not counted multiple times.
We use the Expectation Propagation (EP) algorithm~\cite{Opper2000, Minka2001, Gelman2017} to overcome those difficulties to perform the inference.
The EP algorithm  naturally lends itself to \textit{distributed inference} on partitioned datasets~\mbox{\cite{Gelman2017}}.
Hence we can perform inference on partitions of data, i.e., each scheduled configuration of the HPCs.
In contrast, other techniques for Bayesian inference would require us to explicitly change the inference algorithm depending on the schedule of HPCs and the structure of the FG.
Such changes might not be feasible for all possible schedules or all CPU architectures.
The EP algorithm works by computing an effective region of overlap over our $k$ pieces, i.e., for each piece, we use an approximate prior computed over the other $k-1$ pieces.
The outline of the EP algorithm is illustrated in \cref{alg:ep}.
The algorithm iteratively approximates a target density $f(\cdot)$ (in our case the FG) with a density $g(\cdot)$ that admits the same factorization, and uses a Gaussian \textit{mean field approximation}~\cite{Koller2009}.

\textbf{Training.}
Training is not explicitly required for the proposed BayesPerf model.
The advantage of using Bayesian models like FGs is that training on such models can be reduced to inference on the models' parameters.
At runtime, for each time slice, we compute (infer) a full posterior distribution over the variables (i.e., $E$) and parameters (i.e., $\Theta$) of the FG, and then use maximum likelihood estimation to pick the set of parameters (i.e. $\hat \Theta^{(MLE)}$) that can explain a data trace generated by the system.
    \begin{figure*}[!t]
    \centering
    \includegraphics[width=1.4\columnwidth]{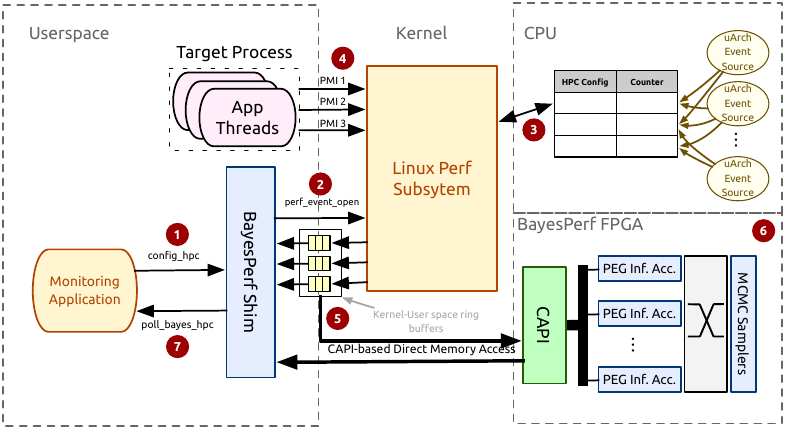}
    \vspace{+0.01in}
    \caption{High-level architecture of the BayesPerf system.}
    \label{fig:bayes_perf_arch}
\end{figure*}

\section{The BayesPerf Implementation}
\label{sec:accelerator}
In this section we describe the software and hardware components in which BayesPerf is deployed.
Further, we describe the architecture and implementation of the BayesPerf accelerator that targets the execution of \cref{alg:ep}.
\cref{fig:bayes_perf_arch} shows the architecture of the BayesPerf system, which works as follows.

\textbf{Setup.} BayesPerf is used by one or more ``monitoring processes/threads'' (labeled ``Monitoring Application'' in \cref{fig:bayes_perf_arch}) to monitor hardware threads of a ``Target Process.''
The BayesPerf user API is identical to the the Linux perf subsystem, and hence any user space program that uses the standard Linux interface can transparently use BayesPerf.
Using this API, the monitoring process registers events of interest (labelled as \circled{1} in \cref{fig:bayes_perf_arch}\footnote{\circled{*} refers to annotations in \cref{fig:bayes_perf_arch} if not otherwise specified.}) with the userspace component of the BayesPerf system, labelled ``BayesPerf Shim.'' The shim represents a userspace driver~\cite{UIO} that replicates the API of the Linux perf subsystem.

\textbf{Linux perf.} The shim registers HPCs on behalf of the user process with the Linux kernel. (labelled as \circled{2}). 
The kernel then manages the scheduling of performance counters onto the CPU (using the scheduling algorithm described in \cref{sec:scheduling_hpcs}). This step is labelled as \circled{3}.
When the target process raises \emph{performance monitoring interrupts} (PMIs; labelled as \circled{4}), the Linux perf subsystem is responsible for reading the corresponding HPC and writing out the sampled value into a ``ring buffer'' (labelled as \circled{5}) that represents a segment of memory that is mapped into the address space of both the shim and the perf subsystem.
The ring buffer represents a FIFO in which new samples are enqueued by the kernel and read from the userspace process.
The ring buffer automatically provides a mechanism for managing backpressure between the shim and kernel as new samples are dropped if the ring buffer is full.

\textbf{Interfacing with the Accelerator.} As we will discuss in \cref{sec:setup}, we have prototyped the BayesPerf system on two different architectures: an Intel x86\_64 and an IBM Power9 processor.
The protocol for communication between the software and the BayesPerf accelerator (labelled as \circled{6}; described later) differ for the two architectures.
On the Power9 system, we leverage CAPI 2.0~\cite{CAPI2015}, a protocol that extends the processor's cache coherence protocol to PCIe-attached devices.
In that case, as the accelerator can directly access the host memory, it can consume samples enqueued onto the ring buffer by the kernel (labelled as \circled{5}).
It does so by snooping on cache invalidation messages for the cache lines corresponding to the ring buffer.
Similarly, outputs of the accelerator are directly written back to the shim's virtual memory space.
For Intel systems, the accelerator uses the base PCIe protocol and IOMMU-mediated PCIe DMA to read HPC samples and write the computed posterior distributions.
Here, the shim must actively poll writes from the kernel to the ring buffer, and once the write has been made, initiate transfer of the samples to the FPGA.
Similarly, the shim polls for interrupts from the accelerator that signify completion of computation, and initiates DMA transfers for the results.
This added software interaction adds some latency overhead to the entire computation.

\textbf{Polling Results.}
Finally, the monitoring application reads (polls) the results of the posterior computation in BayesPerf (labelled as \circled{7}) from ring buffers in the BayesPerf shim.
These reads are always reads to the host memory of the CPU and do not need to initiate DMA requests with the accelerator.
This design is able to mask almost all the latency that is added because of the added computation in BayesPerf (see \cref{fig:overhead}).


\textbf{Multi-Threaded Applications.}
OS-level monitoring contexts, like processes or threads are dealt with at the level of BayesPerf shim.
Hence, when an OS context switch occurs, the memory references of the \texttt{perf} ring buffers are changed by setting configuration registers on the accelerator using MMIO.
When the new references are written, the accelerator begins pulling data from a different buffer in memory.
As a result, the accelerator can be shared across threads that are concurrently executing on the host CPU.

\begin{figure}[!bp]
    \centering
    \includegraphics[width=\columnwidth]{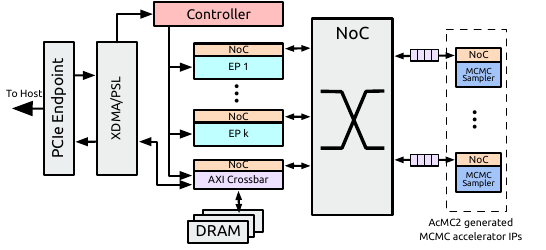}
    \caption{Architecture of the BayesPerf accelerator.}
    \label{fig:accelerator}
\end{figure}

\textbf{The Accelerator.}
\cref{fig:accelerator} illustrates the architecture of the BayesPerf accelerator.
The accelerator exploits parallelism in the structure of \cref{alg:ep} in two ways.
First, we execute posterior inference on each of the $k$ time-slices in parallel (recall \cref{sec:inference}).
These parallel execution engines are labeled as ``EP 1'' through ``EP k'' and ``Controller'' in \cref{fig:accelerator}.
The EPs execute lines 3--6 of \cref{alg:ep} in parallel, and communicate their results to the global controller, which synchronously updates $g(\theta)$ and dispatches the new value to the idle EP.
The values of the measurements from the HPCs (i.e., inputs) as well as the latest values of $g(\theta)$ are stored in the on board DRAM.
Our target FPGA board (which we will describe in \cref{sec:setup}) supports 4 channels of 16GB LPDDR4 memory each.
The input data and the current values of $g(\theta)$ (which together comprise $\sim 100$ MB of data) are replicated across those modules to allow concurrent reads from different EPs to progress simultaneously.

The second level of parallelism exploited by the model is in the computation of MCMC inference in each of the EPs.
Those are represented by the ``MCMC Sampler'' blocks in \cref{fig:accelerator}.
They execute line 4 of \cref{alg:ep} in parallel, by using MCMC to estimate $\Pr(y_k | \theta)$ (i.e., the likelihood that the data $y_k$ is drawn from the local approximator $g_k$).
Here we leverage our prior work, AcMC\textsuperscript{2}\cite{Banerjee_ASPLOS2019}, a high-level synthesis compiler for MCMC applications, to generate IPs that can generate samples from the target distributions of the HPC measurements.
The HPC statistical relationships (i.e., the FG) are fed into the compiler as a probabilistic program, i.e., a program in a domain specific language that can represent statistical dependencies between program variables.
AcMC\textsuperscript{2} then automatically generates efficient uniform random number generators, and automatically synthesizes other statistical constraints in FG.
Instead of using the AcMC\textsuperscript{2}-generated controllers for the MCMC samplers, we use the EPs to directly control the pipelines of MCMC samplers.
That is, 
\begin{enumerate*}[label=(\roman*)]
    \item to set and update configuration parameters like seed values; and
    \item to update the state of the sampler with one which passes the rejection sampling test criteria for each random-walk. 
\end{enumerate*}
Allotment of the samplers to EPs, and all subsequent communication between the EPs and samplers, happen over a network-on-chip (NoC) generated with CONNECT~\cite{Papamichael2012}.
This approach enables us to use samples from previous iterations as starting points for Markov-chain random walks.
This optimization is possible only because we are using MCMC inside an EP algorithm, instead of by itself~\cite{Hoffman2013}.
The NoC uses a butterfly topology to allow communication between EPs and samplers, as well as between the samplers themselves (as is required by AcMC\textsuperscript{2}).
All our experiments use a 16 port NoC, with 4 of those ports being connected to the EPs, and the remaining 12 to the MCMC samplers.
This is the maximal configuration for which we were able to meet timing requirements on the FPGA for a 250 MHz clock.
    \section{Evaluation \& Discussion}
\label{sec:eval}

This section discusses our experimental evaluation of the BayesPerf system and is organized as follows.
First, in \cref{sec:setup}, we describe the experimental setup and explore the performance, power, and area requirements of BayesPerf accelerators when programmed onto an FPGA.
Then, in \cref{sec:measurement} we evaluate the capabilities of the BayesPerf system in correcting measurement errors in HPCs.
Finally, we demonstrate the integration of BayesPerf with ML-based resource management systems to improve their outcomes.

\begin{table}[!t]
    \centering
    \caption{Area \& power for components of the BayesPerf FPGA for the x86\_64 and ppc64 configurations.}
    \resizebox{\columnwidth}{!}{%
    \begin{tabular}{lrrrrrrr}
        \toprule
        \textbf{Component} & \multicolumn{5}{c}{\textbf{Utilization (\%)}} & \multicolumn{2}{c}{\textbf{Power (W)}} \\ \cmidrule(l{2pt}r{2pt}){2-6} \cmidrule(l{2pt}r{2pt}){7-8}
        &\textbf{BRAM} & \textbf{DSP} & \textbf{FF} & \textbf{LUT} & \textbf{URAM} & \textbf{Vivado} & \textbf{Measured} \\
        \midrule
        x86-PCIe & 62 & 78 & 52 & 81 & 58 & 11.2 & 17.2 \\
        ppc64-CAPI & 71 & 66 & 49 & 79 & 58 & 10.5 & 16.1 \\
        \bottomrule
    \end{tabular}}
    \label{tab:fpga_power}
\end{table}

\subsection{Experiment Setup}
\label{sec:setup}
We evaluate BayesPerf on two system configurations:
\begin{enumerate*}[label=(\roman*)]
    \item an IBM AC922 dual-socket Power9 system (which we will refer to as the ``ppc64'' configuration), and
    \item a dual-socket Intel Xeon E5-2695 system (which we will refer to as the ``x86'' configuration).
\end{enumerate*}
Both the systems are populated with two NVIDIA K80 GPUs, a single FDR Infiniband NIC, and a directly attached FPGA board (which we describe below).
Both systems ran Ubuntu 18.04 with kernel version \texttt{v4.15.0}.

\textbf{Accelerator: FPGA.}
The FPGA accelerator was based on the architecture in~\cref{sec:accelerator}.
All experiments were performed on an Alpha-Data ADM-PCIE-9V3 FPGA board (with Xilinx Virtex UltraScale+ VU3P-2 FPGA) clocked at 250 MHz.
For the Power9 systems, the FPGA board was configured to use the CAPI 2.0 interface~\cite{CAPI2015}.
For the x86 configuration, the FPGA board was configured to use PCIe3 x16 along with the Xilinx XRT drivers.
The power and FPGA utilization metrics for the two configurations of the BayesPerf accelerator are listed in \cref{tab:fpga_power}.
In comparison to a 100W TDP of the Intel processor and a 190W TDP Power9 processor, the FPGA performs 5.8\texttimes{} and 11.8\texttimes{} better, respectively, in terms of power consumption.
The BayesPerf-ppc64 FPGA read latency is shown in \cref{fig:overhead}.
We observe that a single HPC read using the CPU implementation of BayesPerf has approximately $9\times$ longer latency than native polling of the HPC.
However, when the accelerator is being used, BayesPerf adds less than 2\% overhead in read latency compared to the native solution.
Compared to the BayesPerf-ppc64 implementation that uses CAPI, the BayesPerf-x86  has on average 15.8\% larger latency.
We can attribute that slowdown to the requirement that a userspace driver actively initiates DMA transfers to the FPGA accelerator, whereas the CAPI configuration snoops for cache invalidation messages.

\subsection{Error Reduction Due to BayesPerf}
\label{sec:measurement}

To demonstrate the efficacy of BayesPerf in correcting HPC measurement errors, we employed the 29 workloads from the HiBench suite~\cite{Hibench}, which span microbenchmarks, machine learning, SQL, web search, graph analytics, and streaming applications.
They represent real-world application workloads used in a cloud environment.
We used the two machines in our experiment to simulate a cluster. Each of the machines hosted 32 workers, and the Spark master was deployed on the x86 node.
We measured 10 derived events for each of the microarchitectures, where each derived event corresponded to a group of HPCs to be measured and aggregated using a mathematical relationship.
We do not detail the events here for lack of space.
The metadata corresponding to the events for the x86 configuration can be found in the Linux kernel source tree~\cite{Kernel_perf} for both the x86 and ppc64 configurations.
In both cases, we measured all HPCs corresponding to the first 10 metrics.

\textbf{Baselines.}
We use three baselines for comparison.
First, we use Linux's inbuilt correction mechanism that uses enabled time and total time (recall from \cref{sec:ml_model}) to correct for measurement errors.
This is the most realistic baseline for users who would use the default configuration available in Linux.
Second, we use a variance reduction technique called CounterMiner~\cite{Lv2018} (CM), a state of the art HPC correction technique used in profiling analysis.
Note that CM was originally meant to be used for offline analysis.
As we will show in the remainder of this section, this requirement manifests as low average correction accuracy, with large variance, when used for online corrections.
Third, we use the online technique by Weaver et. al.~\cite{Weaver2008} (referred to as ``WM+Pin'') for correcting instruction counts in x86 processors.
WM+Pin only corrects the number of instructions executed and was originally meant to correct core performance metrics like IPC or CPI.
Further, it requires intercepting instructions through Pin~\cite{luk2005pin} to collect opcodes for every dynamic instruction. This causes performance degradation of up to 198.2\texttimes{} across our benchmarks.

\begin{figure*}[!t]
    \centering
    \includegraphics[width=\textwidth]{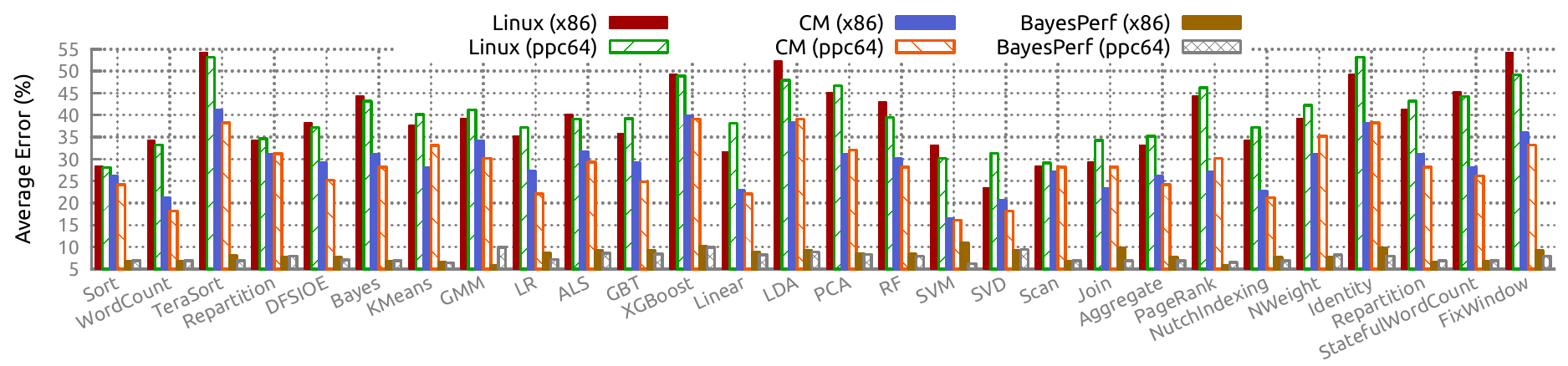}
    \caption{Error in performance counter measurements across the HiBench benchmarks.}
    \label{fig:error_in_bench}
\end{figure*}

\begin{figure*}[!t]
    \centering
    \includegraphics[width=\textwidth]{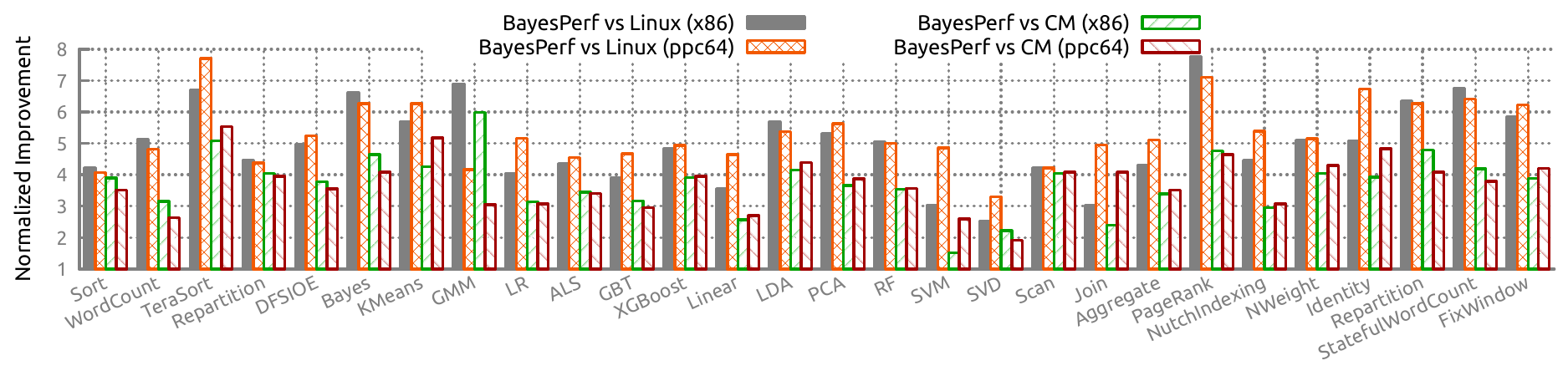}
    \caption{Normalized improvement in performance counter error measurements across the HiBench benchmarks.}
    \label{fig:error_in_bench_norm}
\end{figure*}

\textbf{Error Correction.}
\cref{fig:error_in_bench} shows the significant improvement in measurement values compared to the baseline.
The average error across all benchmarks dropped from 39.25\% and 40.1\% for the ``Linux (x86)'' and ``Linux (ppc64),'' respectively, to 8.06\% (i.e., 4.87\texttimes{}=$\nicefrac{39.25\%}{8.06\%}$) and 7.6\% (i.e., 5.28\texttimes{}=$\nicefrac{40.1\%}{7.6\%}$).
Similarly, when ``BayesPerf (x86)'' and ``BayesPerf (ppc64)'' are compared to ``CM (x86)'' and ``CM (ppc64),'' the average error dropped by 3.63\texttimes{} (=$\nicefrac{29.28\%}{8.06\%}$) and 3.73\texttimes{} ($=\nicefrac{28.31\%}{7.6\%}$), respectively.
Similar improvements were observed in the CM configuration.
That corresponds to a nearly 40\% improvement in the quality of the result of the ppc64 configuration.
The normalized improvement in average error for each of the benchmark applications when using BayesPerf, compared to the two baselines is shown \cref{fig:error_in_bench_norm}.
Recall from \cref{sec:overview}, that error in measurement is computed as the similarity between two time series sequences of performance counter samples~\cite{Berndt1994}.
In the case of the BayesPerf counters, we used a maximum likelihood estimator to provide the most likely value of the performance counter at a point in time.
We normalize the similarity scores using an average similarity score between two runs of the application, where the HPCs were measured with polling.
That way, we could correct for any OS-based nondeterminism in the result.
Just like in \cref{sec:background}, where the magnitude of the error is a comparison between ``polling'' mode and ``sampling'' under Linux and CM (see \cref{fig:error_in_bench}).

\begin{figure}[!bp]
    \centering
    \includegraphics[width=\columnwidth]{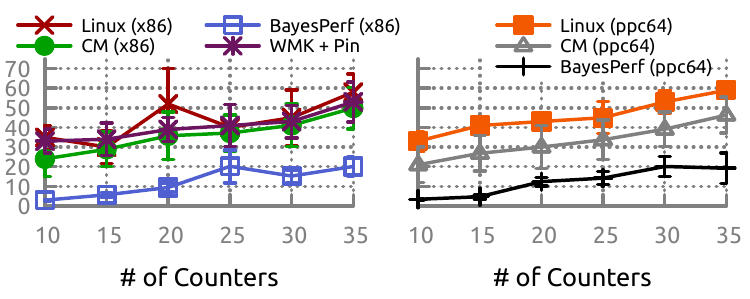}
    \caption{Scaling errors with the number of events sampled.}
    \label{fig:correction_scaling}
\end{figure}

\textbf{Scaling.}
\cref{fig:correction_scaling} shows the scaling behavior of the BayesPerf method with increasing numbers of counters for the ``KMeans'' workload in the HiBench suite.
We observe that BayesPerf consistently reduced error by as much as $34\%$ as the number of counters scaled up from 10 to 35 (for Linux).
Further, WM+Pin performs worse than CM as it only corrects instruction counts.
This justifies our choice of using CM as the main baseline for the evaluation.
Interestingly we find that floating point initialization, which is a major source of errors in~\cite{Weaver2008}, doesn't result in overcounts, indicating that the issue is resolved in modern CPUs.

\textbf{Latency Overhead.}
Since BayesPerf, performs significantly more compute than either Linux or the CM configurations, it is expected to be a significantly higher latency.
Recall from \cref{fig:overhead} that the difference in latency between BayesPerf (when implemented in software) and the Linux correction is nearly 9\texttimes{}.
The BayesPerf accelerator is designed to mitigate the effects of this increased latency.
Again, from \cref{fig:overhead}, we see that it successfully does so, reducing the 9\texttimes{} difference to 2\%.
This is on par with native HPC reads using \texttt{rdpmc} as well as kernel-assisted HPC reads.

\subsection{Case Study: BayesPerf in Feedback Loop}
\label{sec:scheduling}
The core value of the BayesPerf approach in terms of it's error correction capability has been demonstrated in the previous section.
Here we demonstrate the downstream value of BayesPerf to applications that use HPCs as inputs to control system resources.
Examples of such applications include online performance hotspot identification (e.g.,~\cite{Gan2019}), userspace or runtime-level scheduling (e.g.,~\cite{Delimitrou2013, Banerjee_Symphony, Zellweger2016, Giceva2014, Baumann2009}), and power and energy management (e.g.,~\cite{Pothukuchi2018, Pothukuchi2019, Tarsa2019, Ding2019}), as well as attack detectors and system integrity monitors~\cite{Das2019}.
Further they often use as many as 45 HPCs in the case of \cite{10.1145/2254756.2254791, Giceva2014, Banerjee_Symphony}.

\begin{figure}[!t]
    \centering
    \includegraphics[width=\columnwidth]{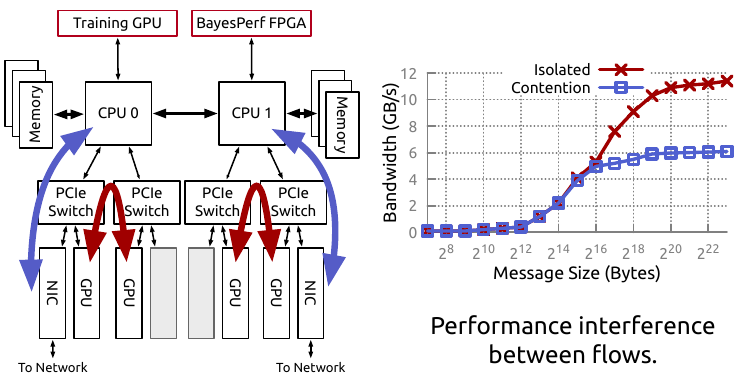}
    \caption{Topology of test system in \cref{sec:scheduling} as well as the effect of the resource contention.}
    \label{fig:topo}
\end{figure}

\textbf{The Problem.}
We now look at a situation in which BayesPerf measurements can be integrated into higher-level decision-making frameworks to perform resource management decisions.
In this part of the experiment, we used HPC measurements to augment an Apache Spark Executor~\cite{Zaharia2016} that needed to run a distributed shuffle operation (which is part of the HiBench TeraSort benchmark~\cite{Hibench}).
\cref{fig:topo} illustrates the rich dynamic information that can be extracted from HPC measurements, and how they can be used in higher-level controllers.
Consider the case of a PCIe interconnect which is populated with NIC and GPU devices.
Here, the Spark executor uses two GPUs to perform a halo exchange (for training a deep neural network).
\cref{fig:topo} shows the performance (in this case, bandwidth) of the exchange as ``isolated'' performance.
If, at the same time, the application were to perform a distributed shuffle (across nodes in a cluster) using the NIC,
we would observe that the original GPU-to-GPU communication is affected because of PCIe bandwidth contention at shared links.
That phenomenon is shown as ``contention'' performance in \cref{fig:topo}, and it can cause as much as a 0--1.8\texttimes{} slowdown, depending on the size of the PCIe transactions.
Online bandwidth and transaction size monitoring (which is enabled by HPCs) can be used by a higher-level software framework to optimally schedule such transfers, so that the performance impact of shared resource contention is minimized.
While the example is simple, it illustrates how errors in measurements can affect the ML algorithm, and hence the overall system performance.

We use two ML-based scheduling algorithms broadly based on those presented in~\cite{Delimitrou2013} and in our prior work~\cite{Banerjee_Symphony}.
The first used collaborative filtering as the core ML algorithm, and the second used deep reinforcement learning.
The goal of our ML-based scheduler was to decide which of the two NICs it would use to perform the shuffle operation, given that the GPUs were communicating with each other and contending for PCIe bandwidth.
We simulated the GPU communication by using Tensorflow to train YoloNet on the ImageNet dataset.

\textbf{The Models.}
The goal of this case study was to show the sensitivity of ML models to errors in their inputs (especially coming from HPCs).
The inputs to the models included:
\begin{enumerate*}[label=(\alph*)]
    \item sampled HPC measurements corresponding to the numbers of allocating, full, partial, and non-snoop writes,
    \item sampled HPC measurements corresponding to demand code reads and partial/MMIO reads,
    \item DRAM Channel bandwidth utilization,
    \item memory-bus bandwidth utilization, and
    \item the size of data to be shuffled (in or out), and the NUMA node on which the data would be resident.
\end{enumerate*}
Note that all of the above are derived events, computation of which required us to capture 32 unique HPC events.
Out of which, 12 were collected for each physical core (i.e., used 432 HPCs = 12 events \texttimes{} 18 cores \texttimes{} 2 sockets), and 20 were off-core events being collected per-socket (i.e., used 40 HPCs = 20 events \texttimes{} 2 sockets).

The first model, used collaborative filtering to impute values of application performance (in this case throughput) with data coming in from the inputs above, as well as data from training workloads of the SparkBench suite in HiBench.
It is based on the technique presented in~\cite{Delimitrou2013}.
The second model used a straightforward neural network: a 4-layer, fully connected ReLU-activated neural network with 36 neurons in layer 1, 16 neurons in each of layers 2 \& 3, and 2 neurons in the last layer.
The two neurons in the last layer chose between the two NICs that were decided between as part of this task.
The model was trained with actor-critic reinforcement learning based on the approach described in~\cite{Banerjee_Symphony}.
The loss function used for training the model minimized the total time taken to complete the shuffle.
The model was trained on the HiBench benchmark suite without the TeraSort benchmark, and then evaluated using the TeraSort benchmark.
When BayesPerf was used, the MLE estimate from the posterior distribution of the HPC was passed into the network.
The GPU marked ``Training GPU'' was used to perform the collaborative filtering and reinforcement learning as well as runtime inference on the system.
It did not contend for the same PCIe resources as the workloads that was being scheduled GPUs.

\begin{figure}[!t]
    \centering
    \includegraphics[width=\columnwidth]{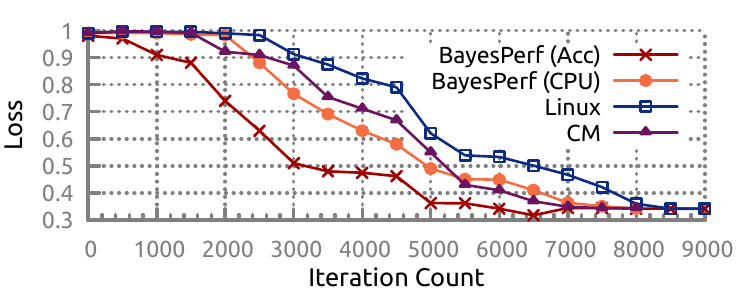}
    \caption{Decrease in training time due to BayesPerf.}
    \label{fig:training_time}
\end{figure}

\textbf{Implementation Details: Training.}
Recall from~\cref{sec:ml_model} that the BayesPerf model in itself does not require training.
However, the two models described above require training.
The model from~\cite{Banerjee_Symphony} learns by reinforcement.
Hence, it does not have specific training and testing phases. The net epochs of data used to train the model are shown in~\cref{fig:training_time}.
For the model in~\cite{Delimitrou2013}, which has specific training and test datasets, we calibrate against bias by using threefold cross-validation (i.e., across applications in~\cref{fig:error_in_bench}).

\textbf{Implementation Details: Hyperparameters.}
The hyperparameters used in the model are taken directly from~\cite{Banerjee_Symphony} and~\cite{Delimitrou2013}.
These parameters include learning rate, LSTM-unroll-length, and epoch lengths, among others.
In addition, we follow the procedure set out in~\cite{Delimitrou2013} to determine the optimal value of sparsity.
We sweep over the range between 30\% and 80\%.
All results in this paper uses the optimal (found from our sweep) value of 75\% sparsity.

\textbf{Results.}
We compare the results of using the above model with BayesPerf and without, using two metrics.

\textbf{Results: Training Time.}
The collaborative filtering model does not have an explicit training phase. For the deep learning model, \cref{fig:training_time} illustrates the difference in training time when error-corrected measurements are used.
In the figure, the loss is normalized using the time taken to run the same shuffle operation in a completely isolated system.
We observed a nearly 37\% reduction in the number of iterations before convergence.
Each training iteration in \cref{fig:training_time} takes 63s; therefore, the overall saving of 37\% corresponds to \textasciitilde{}52 hr.
The reason for the reduction is apparent: a 40\% error in the inputs of the neural network is slowing down the optimization process.
Moreover, we observe that the time to convergence is effected by
\begin{enumerate*}[label=(\alph*)]
    \item the magnitude of error reduction, as seen by the difference between the Linux--CM (12.5\% decrease)  and --BayesPerf (37\% decrease) configurations; and
    \item the timeliness of the error reduction, as seen by the difference between the CPU and accelerated versions of BayesPerf (28.5\% decrease).
\end{enumerate*}

\textbf{Results: Decision Quality.}
We observe that use of the ML-based scheduler (i.e., that makes Spark PCIe aware) leads to a $15.1 \pm 2.2\%$ and $22.3 \pm 7.9\%$ improvement in average shuffle completion time for the two models respectively.
Addition of BayesPerf to the model results in a further $8.7 \pm 0.9\%$ and $19 \pm 3.4\%$ reduction in average shuffle latency, respectively.

    \section{Related Work}
\label{sec:related_work}

\textbf{Error Correction in HPCs}
Measurement errors due to sampling in HPCs have been observed and reported on for the past decade~\cite{Ammons1997, Zellweger2016, Weaver2008, Mytkowicz2007, Dimakopoulou2016, Lv2018, Weaver2013}.
Methods for correction of sampled HPC values can be broadly grouped into two separate approaches.
The first group of methods artificially imputes data in the collected samples by interpolating between two sampled events using linear or piece-wise linear interpolation (e.g.,~\cite{Kernel_perf}).
The advantage of such interpolation methods is that they can be run in real time: however, they might not provide good imputations~\cite{Zellweger2016}.
The second group of methods correct measurements by dropping outlier values, instead of by adding new interpolated values.
Such methods are at the other extreme: they cannot be run in real time, as they need the entire trace of an application before providing corrections.
For example, Lv et al.~\cite{Lv2018} use the Gumbel test for outlier detection, and Neill et. al.~\cite{Neill2017} use fork-join aware agglomorative clustering to remove outlier points.
These methods are not suitable for dynamic control situations that need online HPC correction.
Further, the core statistical technique used by these variance reduction approaches assume that the underlying distribution of the data remains unchanged, however, most workload exhibit distinct stages where workload behavior and thus the underlying distribution of the HPCs will change.

In contrast to those techniques, BayesPerf corrects measurements by using statistical relationships between events.
For well-documented processors, such relationships can be known ahead of time, and the entire correction algorithm can be executed without any need to pre-collect data.
The BayesPerf system (with its accelerator) allows nearly native latency access to the corrected HPCs, thereby enabling their use in dynamic control processes.

\textbf{Using HPCs in Control.}
Several recent papers have explored the use of HPCs to perform higher-level resource management problems.
Examples include online performance hotspot identification (e.g.,~\cite{Gan2019}), userspace or runtime-level scheduling (e.g.,~\cite{Delimitrou2013, Banerjee_Symphony, Zellweger2016, Giceva2014, Baumann2009, Qiu2020, Chen2020}), power and energy management (e.g.,~\cite{Pothukuchi2018, Pothukuchi2019, Tarsa2019, Ding2019}), and attack detectors and system integrity monitors~\cite{Das2019}.
Most of the methods mentioned above do not explicitly use any techniques to correct for errors in HPC measurements.
Further, while it is not impossible that some of the ML techniques can inherently correct for HPC errors, there are no guarantees that it does so.
    \section{Conclusion}
\label{sec:conclusion}

It is crucial to have reliable instrumentation/measurement in commercial CPUs, as exemplified by the inclusion of the \texttt{PEBS} (precision event-based sampling) and \texttt{LBR} (last branch record) technologies in modern Intel processors. However, as we showed in this paper, such technology alone falls short of correcting errors in the values of HPCs accrued because of nondeterminism and sampling artifacts.
This paper presented the design and evaluation of BayesPerf, an ML model and associated accelerator that allows for correction of noisy HPC measurements, reducing the average error in HPC measurements from 42.11\% to 7.8\% when events are being multiplexed.
BayesPerf is the first step in realizing a general-purpose HPC-error-correction system for real x86 and ppc64 systems today and potentially for future processors.
We believe it will form the basis for performing large-scale measurement/characterization studies that use HPC data (i.e., offline analysis), but also enable a slew of applications that can use the HPC data to make control-decisions in a computer system (i.e., online analysis).

\begin{acks}
    We thank the ASPLOS reviewers and our shepherd, Alexandre Passos, for their valuable comments that improved the paper.
    We appreciate S. Lumetta, W-M. Hwu, J. Xiong, and J. Applequist for their insightful discussion and comments on the early drafts of this manuscript.
    This work is partially supported by the National Science Foundation (NSF) under grant Nos. CNS 13-37732, CNS 16-24790, and CCF 20-29049; by the IBM-ILLINOIS Center for Cognitive Computing Systems Research (C3SR), a research collaboration that is part of the IBM AI Horizon Network; and by IBM, Intel, and Xilinx through equipment donations.
    Any opinions, findings, and conclusions or recommendations expressed in this material are those of the authors and do not necessarily reflect the views of the NSF, IBM, Intel, or, Xilinx.
    Saurabh Jha is supported by a 2020 IBM PhD fellowship.
\end{acks}

    \bibliographystyle{ACM-Reference-Format}
    \bibliography{ms}
\end{document}